\documentclass[reprint,superscriptaddress,aps,twocolumn,showpacs]{revtex4-2}
\usepackage{amssymb}
\usepackage{amsmath}
\usepackage{graphicx}
\usepackage{multirow}
\usepackage{subcaption}
\allowdisplaybreaks
\usepackage{physics}
\usepackage{hyperref}
\begin{document}


\title{Group Theory Analysis of $P_c$ Resonances in Molecular Picture}

\author{K. Phumphan}
\email{k.phoompan393@hotmail.com}
\address{School of Physics and Center of Excellence in High Energy Physics and Astrophysics, Suranaree University of Technology, Nakhon Ratchasima 30000, Thailand}
\address{Department of Physics, National Cheng Kung University, Tainan, 70101, Taiwan}
\author{K. Xu}
\email{gxukai1123@gmail.com}
\address{School of Physics and Center of Excellence in High Energy Physics and Astrophysics, Suranaree University of Technology, Nakhon Ratchasima 30000, Thailand}
\author{W. Ruangyoo}
\author{C. C. Chen}
\address{School of Physics and Center of Excellence in High Energy Physics and Astrophysics, Suranaree University of Technology, Nakhon Ratchasima 30000, Thailand}
\address{Department of Physics, National Cheng Kung University, Tainan, 70101, Taiwan}
\author{A. Limphirat}
\email{ayut@g.sut.ac.th}
\author{Y. Yan}
\email{yupeng@g.sut.ac.th}
\address{School of Physics and Center of Excellence in High Energy Physics and Astrophysics, Suranaree University of Technology, Nakhon Ratchasima 30000, Thailand}
\date{\today}

\begin{abstract}
Hidden-charm pentaquarks are analysed in Group Theory in the baryon-meson molecule picture within the framework of the constituent quark model. The investigation of strong decays of pentaquark resonances reveals that pentaquark states of all quark configurations can decay through the $N J/\psi$ channel while only five pentaquark states may decay in open-charm modes. The partial decay widths in the $p\eta_c$ channel are in the same order as in the $p J/\psi$ channel for $J^P=1/2^-$ baryon-meson pentaquark states, but the $p\eta_c$ channel is forbidden for $J^P=3/2^-$ states. The $p \eta_c$, $\Sigma_c\bar{D}$ and $\Lambda_c^+\bar{D}$ channels open for $J^P=1/2^-$ $P_c(4440)$ and $P_c(4457)$ while the $\Sigma_c^*\bar{D}$ channel opens for $J^P=3/2^-$ $P_c(4440)$ and $P_c(4457)$. For  $P_c(4312)$, due to the mass threshold of $\Sigma_c\bar{D}$ and $\Sigma_c^*\bar{D}$ higher than 4312 MeV, the $p \eta_c$ and $\Lambda_c^+\bar{D}$ channels are open for $J^P=1/2^-$ but no open-charm channel open for $J^P=3/2^-$. We strongly suggest that the spin of the $P_c$ resonances may be determined in experiments by investigating the $p \eta_c$ and open-charm channels.

\end{abstract}

\maketitle

\section{\label{sec1}Introduction}
The discovery of pentaquark states, $P_c(4380)$ and $P_c(4450)$ with the quark content of $uudc\bar{c}$ was first reported by the LHCb Collaboration in 2015 in the process $\Lambda_b^0\rightarrow J/\psi p K^-$ \cite{PhysRevLett.115.072001}. In 2019, LHCb updated the observation of the same process from the combined data of RUN 1 and RUN 2. The data analysis shows a new pentaquark state $P_c(4312)$, and $P_c(4450)$ resolved into two pentaquark states, $P_c(4440)$ and $P_c(4457)$ \cite{PhysRevLett.122.222001}. The $P_c$ states have been studied theoretically in various approaches such as QCD sum rules \cite{PhysRevLett.115.172001,PhysRevD.100.051501,PhysRevD.95.094016,PhysRevD.98.054002,Azizi3,Wang:2018waa,Ozdem:2021ugy,Ulas}, one-boson-exchange model \cite{PhysRevLett.115.132002,PhysRevD.100.011502}, effective Lagrangian approach \cite{PhysRevD.100.114002,PhysRevD.100.014022}, coupled-channel interaction \cite{He:2019rva,Kuang:2020bnk} and others \cite{CHEN20161,LIU2019237,ALI2017123,PhysRevLett.122.242001}.

The $P_c$ states were observed in the $pJ/\psi $ channel, which does not necessarily mean that the $pJ/\psi $ mode is their main decay channel. A $qqqc\bar{c}$ system as a baryon-meson molecule state usually decay dominantly via the open-charm decay modes. In contrast, for a genuine pentaquark state, the dominant decay modes are likely the hidden-charm decay channels \cite{CHEN20161}. Therefore, the inner structure of the $P_c$ states may be revealed by studying the decay patterns.

In this work, we study the strong decay patterns of the hidden-charm pentaquark $qqqc\bar{c}$ for all possible strong decay channels, assuming that the hidden-charm pentaquark states are molecule states of a charmed baryon and anticharm meson. The paper is arranged as follows. The quark configurations and wave functions of all ground hidden-charm pentaquark states are derived by applying the group theory \cite{kaifirstpaper,YAN2012496,doi:10.1142/S2010194514602518,PhysRevC.93.025201,PhysRevD.101.076025} in Sec.~\ref{sec2}. In Sec.~\ref{sec3} we evaluated the partial strong decay width ratios of the $P_c(4312)$ resonances. A summary is given in Sec.~\ref{sec4}.

\section{\label{sec2}WAVE FUNCTION CONSTRUCTION}
In this section we work out the quark configuration and wave function of hidden-charm pentaquarks in the molecule picture of $[qqc]$ baryon and $[q\bar{c}]$ meson. The algebraic structure consists of the color and spin-flavor algebras, $SU(3)_c\otimes SU(6)_{sf}$ where $SU(6)_{sf}=SU(3)_f\otimes SU(2)_s$. The permutation symmetries are characterized by the $S_3$ Young tabloids $[3]$, $[21]$ and $[111]$, and the $S_2$ Young tabloids $[2]$ and $[11]$. The total wave function of pentaquark states must be the color-singlet $[222]$ and may be constructed from the states of the baryon and meson clusters,
\begin{equation}\label{totpc}
\Psi_{[222]_C}(q^2cq\bar{c})=\Psi_{[C_{b}]}(q^2c)\otimes \Psi_{[C_{m}]}(q\bar{c}),
\end{equation}
where $\Psi_{[C_{b}]}(q^2c)$ and $\Psi_{[C_{m}]}(q\bar{c})$ are the total wave functions of the baryon and meson clusters, respectively,  with $[C_{b}]$ and $[C_{m}]$ being the color configurations. The color-singlet $[222]$ of the molecule state may be constructed in two ways: the  baryon-meson color singlet-singlet and the color octet-octet configurations. The color part of the pentaquark wave function for the color singlet-singlet configuration is
\begin{equation}\label{col1}
\begin{split}
\psi^c_{[222]}(q^2cq\bar{c})=&~\psi^c_{[111]}(q^2c)\otimes\psi^c_{[111]}(q\bar{c})\\
\psi^c_{[222]}(q^2cq\bar{c})=&~\frac{1}{3\sqrt{2}}(rgb-grb+gbr-rbg\\
&+brg-bgr)(r\bar{r}+g\bar{g}+b\bar{b}),
\end{split}
\end{equation}
where $[111]$ represent for the color singlets of both the baryon and meson clusters. For the color octet-octet configuration, the color wave function of the molecule state takes the form,
\begin{equation}\label{col8}
\psi^{c}_{[222]}=\frac{1}{\sqrt{8}}\sum_j \psi^c_{[21]_i}(q^2c)_j\otimes \psi^c_{[21]}(q\bar{c})_j,
\end{equation}
where $[21]$ represent for the color octets of the baryon and meson clusters, the summation $j$ is over all the color octet states. The details of the wave functions are given in Appendix~\ref{Apcol8}.

For the baryon cluster, the configuration of the spatial, flavor and spin parts can be derived by following the fact that the total wave function of the baryon cluster is antisymmetric for the light quarks in the cluster. The total wave function of the baryon cluster can be written in terms of the color and spatial-spin-flavor wave functions,
\begin{equation}
\begin{split}
&\Psi(q^2c)_{singlet}=\psi^c_{[111]}\left(\psi^{osf}_{[2]}(q^2)\otimes \psi^{osf}(c)\right),\\
&\Psi(q^2c)_{octet_\lambda}=\psi^c_{[21]_\lambda}\left(\psi^{osf}_{[11]}(q^2)\otimes\psi^{sf}(c)\right),\\
&\Psi(q^2c)_{octet_\rho}=\psi^c_{[21]_\rho}\left(\psi^{osf}_{[2]}(q^2)\otimes\psi^{sf}(c)\right),\\
\end{split}
\end{equation}
with
\begin{equation}
\begin{split}
&\psi^{osf}_{[2]}=\left\lbrace \psi^o_{[2]}\psi^{sf}_{[2]}, \psi^o_{[11]}\psi^{sf}_{[11]} \right\rbrace,\\
&\psi^{osf}_{[11]}=\left\lbrace \psi^o_{[2]}\psi^{sf}_{[11]}, \psi^o_{[11]}\psi^{sf}_{[2]}
 \right\rbrace,
\end{split}
\end{equation} where $\psi^{osf}$, $\psi^{sf}$, $\psi^c$, $\psi^o$, $\psi^s$ and $\psi^f$ are the spatial-spin-flavor, spin-flavor, color, spin, flavor wave functions, respectively. $\lambda$ and $\rho$ stand for the mixed-symmetric and mixed-antisymmetric states of the $[21]$ representation, respectively. For the ground state of pentaquarks, the spatial wave function is symmetric, and thus the total wave functions of the baryon cluster takes the form,
\begin{equation}\label{b2c}
\begin{split}
&\Psi(q^2c)_{singlet}=\psi^c_{[111]}\psi^o_{[3]}\left(\psi^{sf}_{[2]}(q^2)\otimes\psi^{sf}(c)\right),\\
&\Psi(q^2c)_{octet_\lambda}=\psi^c_{[21]_\lambda}\psi^o_{[3]}\left(\psi^{sf}_{[11]}(q^2)\otimes\psi^{sf}(c)\right),\\
&\Psi(q^2c)_{octet_\rho}=\psi^c_{[21]_\rho}\psi^o_{[3]}\left(\psi^{sf}_{[2]}(q^2)\otimes\psi^{sf}(c)\right),\\
\end{split}
\end{equation}
where
\begin{equation}\label{sf211}
\begin{split}
&\psi^{sf}_{[2]}=\left\lbrace  \psi^f_{[2]}\psi^s_{[2]}, \psi^f_{[11]}\psi^s_{[11]}  \right\rbrace,\\
&\psi^{sf}_{[11]}=\left\lbrace  \psi^f_{[2]}\psi^s_{[11]}, \psi^f_{[11]}\psi^s_{[2]}  \right\rbrace.\\
\end{split}
\end{equation}
The products of $\psi^{sf}_{[x]}(q^2)\otimes\psi^{sf}(c)$ in Eq.~(\ref{b2c}) lead to \begin{equation}\label{fb}
\begin{split}
&\psi^{sf}_{[2]}(q^2c)=\left\lbrace  \psi^f_{[2]}\psi^s_{[3]},  \psi^f_{[2]}\psi^s_{[21]_\lambda}, \psi^f_{[11]}\psi^s_{[21]_\rho}  \right\rbrace,\\
&\psi^{sf}_{[11]}(q^2c)=\left\lbrace \psi^f_{[2]}\psi^s_{[21]\rho}, \psi^f_{[11]}\psi^s_{[3]}, \psi^f_{[11]}\psi^s_{[21]_\lambda} \right\rbrace ,
\end{split}
\end{equation}
with $\psi^f_{[x]}\equiv\psi^f_{[x]}(q^2c)$ being the flavor part with $[x]$ representing for the flavor configuration of $q^2$, and $\psi^s_{[y]}\equiv\psi^s_{[y]}(q^2c)$ being the spin part with $[y]$ being the spin configurations of the baryon cluster.

The total wave functions of the meson cluster can be written as \begin{equation}\label{m2c}
\begin{split}
&\Psi(q\bar{c})_{singlet}=\psi^c_{[111]}\psi^o_{[2]}\psi^{f}(q\bar{c})\psi^s_{[z]},\\
&\Psi(q\bar{c})_{octet}=\psi^c_{[21]}\psi^o_{[2]}\psi^{f}(q\bar{c})\psi^s_{[z]},
\end{split}
\end{equation}
with $\psi^{s}_{[z]}=\left\lbrace \psi^s_{[2]},\psi^{s}_{[11]}\right\rbrace $ corresponding to spin $S=1$ and $S=0$ of the meson cluster, respectively. The total wave function of pentaquarks is derived by substituting the wave functions of the baryon and meson clusters in Eqs.~(\ref{b2c}) and~(\ref{m2c}) into Eq.~(\ref{totpc}). Finally, we derive 18 quark configurations and totally 47 possible molecule states of ground state pentaquarks, as listed explicitly in Table~\ref{pccon}. 
\begin{table*}[t]
\centering
\caption{$P_c$ configurations and states.\label{pccon}}
\begin{ruledtabular}
\begin{tabular}{l l c l c c}
Pentaquark configuration&$\Psi(q^2c)$& $\otimes$&$\Psi(q\bar{c})$ &Isospin& Spin\\
\hline
$\Psi_{[111]_C[2]_F[3][2]_S}$&$\psi_{[111]}\phi_{[2]}\chi_{[3]}$&$\otimes$ & $\psi_{[111]}\phi(q\bar{c})\chi_{[2]}$&$\frac{3}{2},\frac{1}{2}$ &$\frac{5}{2},\frac{3}{2},\frac{1}{2}$\\
$\Psi_{[111]_C[2]_F[3][11]_S}$&$\psi_{[111]}\phi_{[2]}\chi_{[3]}$&$\otimes$ & $\psi_{[111]}\phi(q\bar{c})\chi_{[11]}$&$\frac{3}{2},\frac{1}{2}$ &$\frac{3}{2}$\\
$\Psi_{[111]_C[2]_F[21][2]_S}$&$\psi_{[111]}\phi_{[2]}\chi_{[21]_\lambda}$&$\otimes$ & $\psi_{[111]}\phi(q\bar{c})\chi_{[2]}$&$\frac{3}{2},\frac{1}{2}$ &$\frac{3}{2},\frac{1}{2}$\\
$\Psi_{[111]_C[2]_F[21][11]_S}$&$\psi_{[111]}\phi_{[2]}\chi_{[21]_\lambda}$&$\otimes$ & $\psi_{[111]}\phi(q\bar{c})\chi_{[11]}$&$\frac{3}{2},\frac{1}{2}$ &$\frac{1}{2}$\\
$\Psi_{[111]_C[11]_F[21][2]_S}$&$\psi_{[111]}\phi_{[11]}\chi_{[21]_\rho}$&$\otimes$ & $\psi_{[111]}\phi(q\bar{c})\chi_{[2]}$&$\frac{1}{2}$ &$\frac{3}{2},\frac{1}{2}$\\
$\Psi_{[111]_C[11]_F[21][11]_S}$&$\psi_{[111]}\phi_{[11]}\chi_{[21]_\rho}$&$\otimes$ & $\psi_{[111]}\phi(q\bar{c})\chi_{[11]}$&$\frac{1}{2}$ &$\frac{1}{2}$\\
\hline
$\Psi_{[21]^\lambda_C[2]_F[21]^\rho[2]_S}$&$\psi_{[21]_\lambda} \phi_{[2]} \chi_{[21]_\rho}$&$\otimes$&$\psi_{[21]}\phi(q\bar{c})\chi_{[2]}$&$\frac{3}{2},\frac{1}{2}$&$\frac{3}{2},\frac{1}{2}$\\
$\Psi_{[21]^\lambda_C[2]_F[21]^\rho[11]_S}$&$\psi_{[21]_\lambda} \phi_{[2]} \chi_{[21]_\rho}$&$\otimes$&$\psi_{[21]}\phi(q\bar{c})\chi_{[11]}$&$\frac{3}{2},\frac{1}{2}$&$\frac{1}{2}$\\
$\Psi_{[21]^\lambda_C[11]_F[3][2]_S}$&$\psi_{[21]_\lambda} \phi_{[11]} \chi_{[3]}$&$\otimes$&$\psi_{[21]}\phi(q\bar{c})\chi_{[2]}$&$\frac{1}{2}$&$\frac{5}{2},\frac{3}{2},\frac{1}{2}$\\
$\Psi_{[21]^\lambda_C[11]_F[3][11]_S}$&$\psi_{[21]_\lambda} \phi_{[11]} \chi_{[3]}$&$\otimes$&$\psi_{[21]}\phi(q\bar{c})\chi_{[11]}$&$\frac{1}{2}$&$\frac{3}{2}$\\
$\Psi_{[21]^\lambda_C[11]_F[21]^\lambda[2]_S}$&$\psi_{[21]_\lambda} \phi_{[11]} \chi_{[21]_\lambda}$&$\otimes$&$\psi_{[21]}\phi(q\bar{c})\chi_{[2]}$&$\frac{1}{2}$&$\frac{3}{2},\frac{1}{2}$\\
$\Psi_{[21]^\lambda_C[11]_F[21]^\lambda[11]_S}$&$\psi_{[21]_\lambda} \phi_{[11]} \chi_{[21]_\lambda}$&$\otimes$&$\psi_{[21]}\phi(q\bar{c})\chi_{[11]}$&$\frac{1}{2}$&$\frac{1}{2}$\\
\hline
$\Psi_{[21]^\rho_C[2]_F[3][2]_S}$&$\psi_{[21]_\rho} \phi_{[2]} \chi_{[3]}$&$\otimes$&$\psi_{[21]}\phi(q\bar{c})\chi_{[2]}$&$\frac{3}{2},\frac{1}{2}$&$\frac{5}{2},\frac{3}{2},\frac{1}{2}$\\
$\Psi_{[21]^\rho_C[2]_F[3][11]_S}$&$\psi_{[21]_\rho} \phi_{[2]} \chi_{[3]}$&$\otimes$&$\psi_{[21]}\phi(q\bar{c})\chi_{[11]}$&$\frac{3}{2},\frac{1}{2}$&$\frac{3}{2}$\\
$\Psi_{[21]^\rho_C[2]_F[21]^\lambda[2]_S}$&$\psi_{[21]_\rho} \phi_{[2]} \chi_{[21]_\lambda}$&$\otimes$&$\psi_{[21]}\phi(q\bar{c})\chi_{[2]}$&$\frac{3}{2},\frac{1}{2}$&$\frac{3}{2},\frac{1}{2}$\\
$\Psi_{[21]^\rho_C[2]_F[21]^\lambda[11]_S}$&$\psi_{[21]_\rho} \phi_{[2]} \chi_{[21]_\lambda}$&$\otimes$&$\psi_{[21]}\phi(q\bar{c})\chi_{[11]}$&$\frac{3}{2},\frac{1}{2}$&$\frac{1}{2}$\\
$\Psi_{[21]^\rho_C[11]_F[21]^\rho[2]_S}$&$\psi_{[21]_\rho} \phi_{[11]} \chi_{[21]_\rho}$&$\otimes$&$\psi_{[21]}\phi(q\bar{c})\chi_{[2]}$&$\frac{1}{2}$&$\frac{3}{2},\frac{1}{2}$\\
$\Psi_{[21]^\rho_C[11]_F[21]^\rho[11]_S}$&$\psi_{[21]_\rho} \phi_{[11]} \chi_{[21]_\rho}$&$\otimes$&$\psi_{[21]}\phi(q\bar{c})\chi_{[11]}$&$\frac{1}{2}$&$\frac{1}{2}$\\
\end{tabular}
\end{ruledtabular}
\end{table*}
\section{\label{sec3}Transition amplitude \lowercase{and} Decay widths}
\begin{figure}[!t]
\centering
\begin{subfigure}[b]{0.4\textwidth}
\centering
\includegraphics[scale=0.4]{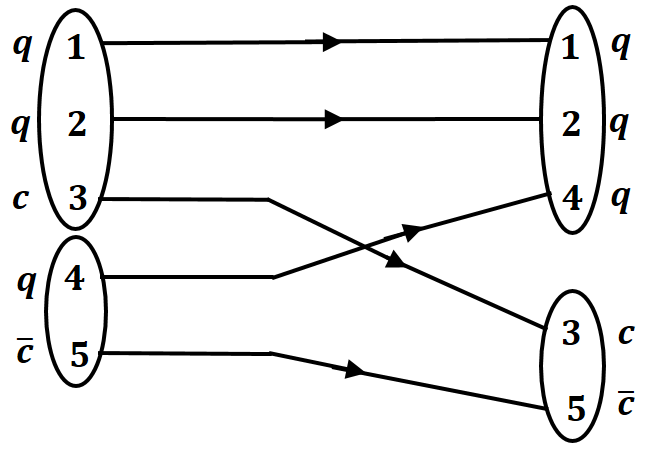}
\caption{\label{da}}
\end{subfigure}
\quad
\begin{subfigure}[b]{0.4\textwidth}
\centering
\includegraphics[scale=0.4]{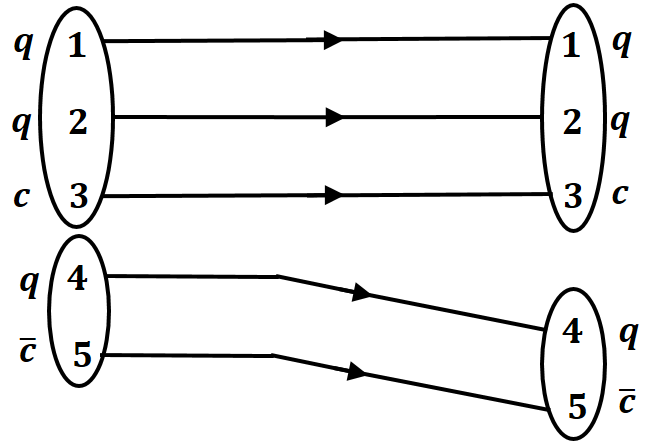}
\caption{\label{db}}
\end{subfigure}
\caption{Quark interchange diagrams for $P_c$ in (a) the hidden-charm decay modes and (b) the open-charm decay modes.\label{diagram}}
\end{figure}
In this work, we study the decay of non-strange hidden-charm pentaquarks in the processes shown in the diagrams in Figure~\ref{diagram}. The decay width for two-body decays can be calculated with \cite{osti_4002515}
\begin{equation}
\begin{split}
\Gamma_{a\rightarrow b}=\frac{2\pi}{2S_{a}+1}\sum_{m_a,m_b}\int &d\vec{p}_1 d\vec{p}_2 \delta(E_b-E_a)\\
&\times\delta(\vec{p}_b-\vec{p}_a){|T_{ba}|}^2
\end{split}
\end{equation}
where $\vec{p}_1$ and $\vec{p}_2$ are the momentum of final particles, $E_a$ and $E_b$ are the energy of initial and final states, $S_{a}$ and $m_a$ are the spin of the initial state and its third component, and $T_{ba}$ is the transition amplitude. The transition amplitude $T_{ba}$ may be factorized by the color-spin-flavor and spatial parts, and hence
the decay width may be written in the form,
\begin{equation}\label{decaywidth}
\Gamma_{P_c\rightarrow ij}=\frac{2\pi E_{i}E_{j}}{M_{P_c}}f(p^2)\abs{\gamma}^2,
\end{equation}
where $E_{i}$ and $E_j$ are the energies of the final two particles, $M_{P_c}$ is the mass of pentaquark state. $f(p^2)$ is the kinematic phase space factor stemming from the spatial part of the transition amplitude, depending on the final particle momentum $p$ and the spatial wave functions of initial and final states. The transition factor $\gamma$ can be calculated, for example, for the hidden-charm decay modes by
\begin{widetext}
\begin{equation}
\begin{split}
\gamma =& \bra{(q^3)_{[111]_C}\otimes (c\bar{c})_{[111]_C}}\ket{(q^2cq\bar{c})_{[222]}}\\&~\times\bra{\left[ \left(q^2_{I_{2},J_{2}}\otimes q \right)_{I_B,J_B}\otimes\left(c\bar{c}\right)_{I_M,J_M}\right]_{I_{f},J_f}}\ket{\left[\left(q^2_{I_{1},J_1}\otimes c\right)_{I_b,J_b}\otimes \left(q\bar{c}\right)_{I_m,J_m}\right]_{I_{P_c},J_{P_c}}}\\
=&~\gamma_C\delta_{I_1,I_2}\delta_{I_{P_c},I_f}\delta_{J_1,J_2}\delta_{J_{P_c},J_f}\sqrt{(2j_B+1)(2j_M+1)(2j_b+1)(2j_m+1)}\begin{Bmatrix}
j&1/2&j_B\\1/2&1/2&j_M\\j_b&j_m&j_{P_c}
\end{Bmatrix},
\end{split}
\end{equation}
\end{widetext}
where $\{\}$ is the Wigner's 9-j symbol and $\gamma_C$ is the transition factor of color part. All non-vanishing results of the color-spin-flavor factor $\gamma$ are shown in Table~\ref{csf3}. Note that the states with the color octet $\lambda$ type of the baryon cluster can not proceed in the strong decay in the present model due to the color factor.
\begin{table*}[t]
\centering
\caption{Color-Spin-Flavor factors of all possible channels.\label{csf3}}
\begin{ruledtabular}
\begin{tabular}{c c l c c c c c c c c c c}
$\vb{I}$&$\vb{j}$&$P_c$ configuration &$NJ/\psi$&$N\eta_c$&$\Delta^+J/\psi$&$\Delta^+\eta_c$&$\Sigma_c^*\bar{D}^*$&$\Sigma_c\bar{D}^*$&$\Lambda^+_c\bar{D}^*$&$\Sigma_c^*\bar{D}$&$\Sigma_c\bar{D}$&$\Lambda^+_c\bar{D}$\\
\hline
$\frac{3}{2}$&$\frac{5}{2}$&$\Psi_{[111]_C[2]_F[3][2]_S}$&&&$\frac{1}{3}$&&1&&&&&\\
&&$\Psi_{[21]^\rho_C[2]_F[3][2]_S}$&&&$\frac{2\sqrt{2}}{3}$&&&&&&&\\
\hline
$\frac{3}{2}$&$\frac{3}{2}$&$\Psi_{[111]_C[2]_F[3][2]_S}$&&&$\frac{1}{18}$&$\frac{1}{6}\sqrt{\frac{5}{3}}$&1&&&&&\\
&&$\Psi_{[111]_C[2]_F[21][2]_S}$&&&$\frac{\sqrt{5}}{9}$&$-\frac{1}{3\sqrt{3}}$&&1&&&&\\
&&$\Psi_{[111]_C[2]_F[3][11]_S}$&&&$\frac{1}{6}\sqrt{\frac{5}{3}}$&$\frac{1}{6}$&&&&1&&\\
&&$\Psi_{[21]^\rho_C[2]_F[3][2]_S}$&&&$\frac{\sqrt{2}}{9}$&$\frac{1}{3}\sqrt{\frac{10}{3}}$&&&&&&\\
&&$\Psi_{[21]^\rho_C[2]_F[3][11]_S}$&&&$\frac{1}{3}\sqrt{\frac{10}{3}}$&$\frac{\sqrt{2}}{6}$&&&&&&\\
&&$\Psi_{[21]^\rho_C[2]_F[21][2]_S}$&&&$\frac{2\sqrt{10}}{9}$&$-\frac{2}{6}\sqrt{\frac{2}{3}}$&&&&&&\\
\hline
$\frac{3}{2}$&$\frac{1}{2}$&$\Psi_{[111]_C[2]_F[3][2]_S}$&&&$-\frac{1}{9}$&&1&&&&&\\
&&$\Psi_{[111]_C[2]_F[21][2]_S}$&&&$\frac{\sqrt{2}}{9}$&&&1&&&&\\
&&$\Psi_{[111]_C[2]_F[21][11]_S}$&&&$\frac{1}{3}\sqrt{\frac{2}{3}}$&&&&&&1&\\
&&$\Psi_{[21]^\rho_C[2]_F[3][2]_S}$&&&$-\frac{2\sqrt{2}}{9}$&&&&&&&\\
&&$\Psi_{[21]^\rho_C[2]_F[21][11]_S}$&&&$\frac{4}{3\sqrt{3}}$&&&&&&&\\
&&$\Psi_{[21]^\rho_C[2]_F[21][2]_S}$&&&$\frac{4}{9}$&&&&&&&\\
\hline
$\frac{1}{2}$&$\frac{5}{2}$&$\Psi_{[111]_C[2]_F[3][2]_S}$&&&&&1&&&&&\\
\hline
$\frac{1}{2}$&$\frac{3}{2}$&$\Psi_{[111]_C[2]_F[3][2]_S}$&$\frac{1}{9}\sqrt{\frac{5}{2}}$&&&&1&&&&&\\
&&$\Psi_{[111]_C[2]_F[21][2]_S}$&$\frac{1}{9\sqrt{2}}$&&&&&1&&&&\\
&&$\Psi_{[111]_C[11]_F[21][2]_S}$&$\frac{1}{3\sqrt{2}}$&&&&&&1&&&\\
&&$\Psi_{[111]_C[2]_F[3][11]_S}$&$-\frac{1}{3\sqrt{6}}$&&&&&&&1&&\\
&&$\Psi_{[21]^\rho_C[2]_F[3][2]_S}$&$\frac{2\sqrt{5}}{9}$&&&&&&&&&\\
&&$\Psi_{[21]^\rho_C[2]_F[3][11]_S}$&$-\frac{2}{3\sqrt{3}}$&&&&&&&&&\\
&&$\Psi_{[21]^\rho_C[2]_F[21]^\lambda[2]_S}$&$\frac{2}{9}$&&&&&&&&&\\
&&$\Psi_{[21]^\rho_C[11]_F[21]^\rho[2]_S}$&$\frac{2}{3}$&&&&&&&&&\\
\hline
$\frac{1}{2}$&$\frac{1}{2}$&$\Psi_{[111]_C[2]_F[3][2]_S}$&$\frac{1}{9}$&$\frac{1}{3\sqrt{3}}$&&&1&&&&&\\
&&$\Psi_{[111]_C[2]_F[21][2]_S}$&$\frac{5}{18\sqrt{2}}$&$-\frac{1}{6\sqrt{6}}$&&&&1&&&&\\
&&$\Psi_{[111]_C[11]_F[21][2]_S}$&$-\frac{1}{6\sqrt{2}}$&$\frac{1}{2\sqrt{6}}$&&&&&1&&&\\
&&$\Psi_{[111]_C[2]_F[21][11]_S}$&$-\frac{1}{6\sqrt{6}}$&$\frac{1}{6\sqrt{2}}$&&&&&&&1&\\
&&$\Psi_{[111]_C[11]_F[21][11]_S}$&$\frac{1}{2\sqrt{6}}$&$\frac{1}{6\sqrt{2}}$&&&&&&&&1\\
&&$\Psi_{[21]^\rho_C[2]_F[3][2]_S}$&$\frac{2\sqrt{2}}{9}$&$\frac{2}{3}\sqrt{\frac{2}{3}}$&&&&&&&&\\
&&$\Psi_{[21]^\rho_C[2]_F[21]^\lambda[2]_S}$&$\frac{5}{9}$&$-\frac{1}{3\sqrt{3}}$&&&&&&&&\\
&&$\Psi_{[21]^\rho_C[2]_F[21]^\lambda[11]_S}$&$-\frac{1}{3\sqrt{3}}$&$\frac{1}{3}$&&&&&&&&\\
&&$\Psi_{[21]^\rho_C[11]_F[21]^\rho[2]_S}$&$-\frac{1}{3}$&$\frac{1}{\sqrt{3}}$&&&&&&&&\\
&&$\Psi_{[21]^\rho_C[11]_F[21]^\rho[11]_S}$&$\frac{1}{\sqrt{3}}$&$\frac{1}{3}$&&&&&&&&\\
\end{tabular}
\end{ruledtabular}
\end{table*}

A study of nucleon-antinucleon annihilations to two mesons \cite{DOVER199287}, where the spatial wave functions of particles were approximated in the Gaussian form from MIT bag model, shows that an analytical kinematic factor of overlapping wave functions takes the from as $f(p^2)\approx p\exp{-\alpha p^2 R^2}$, where $R$ is the bag radius of particles and $\alpha$ can be obtained by fitting to experimental data. It is shown in Ref. \cite{DOVER199287} that a phenomenological phase factor can produce the same nucleon-antinucleon annihilation branching ratios at low energies as the analytical kinematic factor. To reduce the the model dependence, we apply the phenomenological form for the phase space factor,
\begin{equation}\label{kpsf}
f(m_i,m_j,\sqrt{s})=p\cdot C\exp{-A(s-s_{ij})^{1/2}}
\end{equation}
where $p$ is the center-of-mass momentum, $s_{ij}=(m_i+m_j)^2$, $C$ is a constant and $A=-1.2\mathrm{GeV^{-1}}$. $C$ and $A$ are assumed to be equal for all decay modes. The kinematical phase-space factor in Eq. (\ref{kpsf}) has been fitted to the cross section of various $p\bar p$ annihilation channels \cite{Vandermeulen:1988hh}.

The partial decay widths of hidden-charm pentaquarks are calculated in Eq.~(\ref{decaywidth}). Listed in Table~\ref{ratio} are the partial decay widths of all possible decay channels, normalized to the decay width of $\Psi_{[111]_C[2]_F[3][2]_S}({\bf J}=3/2)\rightarrow pJ/\psi$ decay channel.
The decay widths of the three narrow $P_c$ resonances are very close, and thus we show in Table~\ref{ratio} only the results with the mass of $P_c(4457)$ as input.

\begin{table}[t]
\centering
\caption{$P_c(4457)$ $I=1/2$ decay widths normalized by $\Gamma(\Psi_{[111]_C[2]_F[3][2]_S}^{J=3/2}\rightarrow pJ/\psi)$. \label{ratio}}
\begin{ruledtabular}
\begin{tabular}{c l c c c c c c c}
$J$&$P_c$ configuration &$p\eta_c$&$pJ/\psi$&$\Sigma_c^* \bar{D}$&$\Sigma_c \bar{D}$&$\Lambda^+_c\bar{D}$&$\Lambda^+_c\bar{D}^*$\\
\hline
\multirow{8}{*}{$\frac{3}{2}$}&$\Psi_{[111]_C[2]_F[3][2]_S}$&&1&&&&&\\
&$\Psi_{[111]_C[2]_F[3][11]_S}$&&0.59&69.63&&&&\\
&$\Psi_{[111]_C[2]_F[21][2]_S}$&&0.19&&&&&\\
&$\Psi_{[111]_C[11]_F[21][2]_S}$&&1.80&&&&66.71&\\
&$\Psi_{[21]^\rho_C[2]_F[3][2]_S}$&&8.00&&&&&\\
&$\Psi_{[21]^\rho_C[2]_F[3][11]_S}$&&4.80&&&&&\\
&$\Psi_{[21]^\rho_C[2]_F[21]^\lambda[2]_S}$&&1.59&&&&&\\
&$\Psi_{[21]^\rho_C[11]_F[21]^\rho[2]_S}$&&14.40&&&&&\\
\hline
\multirow{10}{*}{$\frac{1}{2}$}&$\Psi_{[111]_C[2]_F[3][2]_S}$&1.07&0.39&&&&&\\
&$\Psi_{[111]_C[2]_F[21][2]_S}$&0.13&1.25&&&&&\\
&$\Psi_{[111]_C[2]_F[21][11]_S}$&0.40&0.14&&67.51&&&\\
&$\Psi_{[111]_C[11]_F[21][2]_S}$&1.21&0.44&&&&66.71&\\
&$\Psi_{[111]_C[11]_F[21][11]_S}$&0.40&1.34&&&53.94&&\\
&$\Psi_{[21]^\rho_C[2]_F[3][2]_S}$&8.63&3.20&&&&&\\
&$\Psi_{[21]^\rho_C[2]_F[21]^\lambda[2]_S}$&1.09&10.00&&&&&\\
&$\Psi_{[21]^\rho_C[2]_F[21]^\lambda[11]_S}$&3.23&1.19&&&&&\\
&$\Psi_{[21]^\rho_C[11]_F[21]^\rho[2]_S}$&9.71&3.60&&&&&\\
&$\Psi_{[21]^\rho_C[11]_F[21]^\rho[11]_S}$&3.23&10.80&&&&&\\
\end{tabular}
\end{ruledtabular}
\end{table}

\section{\label{sec4}Discussion}
We have calculated the partial decay widths of the charmonium-like pentaquark states, assuming that the pentaquark resonances are baryon-meson molecules, where 18 quark configurations and totally 47 possible molecule states of ground state pentaquarks are considered. The results in Table~\ref{ratio} show that the baryon-meson pentaquark states of all quark configurations in the model can decay through the $N J/\psi$ channel, but only five pentaquark states may decay through open-charm modes, and the open charm decays, once allowed, dominate over the the hidden-charm channels.

It is found in Table~\ref{ratio} that the $p\eta_c$ channel is forbidden for $J^P=3/2^-$ baryon-meson pentaquark states, but the partial decay widths in the $p\eta_c$ channel are in the same order as in the $p J/\psi$ channel for $J^P=1/2^-$ states. For $P_c(4440)$ and $P_c(4457)$, the $p \eta_c$, $\Sigma_c\bar{D}$ and $\Lambda_c^+\bar{D}$ channels are open only for $J^P=1/2^-$ and the $\Sigma_c^*\bar{D}$ channel is open only for $J^P=3/2^-$. For $P_c(4312)$, due to the mass threshold of $\Sigma_c\bar{D}$ and $\Sigma_c^*\bar{D}$ higher than 4312 MeV, the $p \eta_c$ and $\Lambda_c^+\bar{D}$ channels are open for $J^P=1/2^-$ but no open-charm channel open for $J^P=3/2^-$. We strongly suggest that charmonium-like pentaquarks to be searched in experiments in both the open charm and hidden charm channels.

The significant event rates in the open charm channels are estimated to be in the same order as in the hidden charm channels. Assuming that the process $P_c\rightarrow \Sigma_c\bar D\; ( \Sigma_c\rightarrow \Lambda_c \pi^+,\; \Lambda_c \rightarrow p K^- \pi^+)$ is investigated in experiments, the present work predicts a decay ratio, 0.35 relative to the $p J/\psi$ channel (if $J/\psi$ has been 100\% recovered), where we have used $BR(\Sigma_c\rightarrow \Lambda_c \pi^+)= 100\%$, $BR(\Lambda_c \rightarrow p K^- \pi^+)= 5\%$, and $BR(D^+\rightarrow K^-\pi^+\pi^+)= 10\%$.

\begin{acknowledgments}
This work is supported by Suranaree University of Technology (SUT) and National Cheng Kung University (NCKU). K.P., W.R., and C.C.C. acknowledge support from SUT and NCKU. A.L. and Y.Y. acknowledge support from SUT.

\end{acknowledgments}

\appendix

\section{\label{Apcol8}Color octet wave functions}
The construction of pentaquark wave functions in the molecule picture has been shown in Sec.~\ref{sec2}. The explicit form of the color octet wave functions of the baryon cluster and their conjugates of the meson cluster are shown in Eqs.~(\ref{collambda})-(\ref{colmes}). The color octet $\lambda$ type states of the baryon cluster are
\begin{equation}\label{collambda}
\begin{split}
\psi_{[21]_\lambda}(q^2c)_1=&~\frac{1}{\sqrt{6}}(2rrg-rgr-grr),\\
\psi_{[21]_\lambda}(q^2c)_2=&~\frac{1}{\sqrt{6}}(grg+rgg-2ggr),\\
\psi_{[21]_\lambda}(q^2c)_3=&~\frac{1}{\sqrt{6}}(2rrb-rbr-brr),\\
\psi_{[21]_\lambda}(q^2c)_4=&~\frac{1}{\sqrt{12}}(2rgb+2grb-gbr\\
&-rbg-brg-bgr),\\
\psi_{[21]_\lambda}(q^2c)_5=&~\frac{1}{\sqrt{6}}(2ggb-gbg-bgg),\\
\psi_{[21]_\lambda}(q^2c)_6=&~\frac{1}{\sqrt{6}}(brb+rbb-2bbr),\\
\psi_{[21]_\lambda}(q^2c)_7=&~\frac{1}{\sqrt{6}}(bgb+gbb-2bbg),\\
\psi_{[21]_\lambda}(q^2c)_8=&~\frac{1}{\sqrt{4}}(rbg+brg-bgr-gbr).
\end{split}
\end{equation}

The color octet $\rho$ type states of the baryon cluster are
\begin{equation}\label{colrho}
\begin{split}
\psi_{[21]_\rho}(q^2c)_1=&~\frac{1}{\sqrt{2}}(rgr-grr),\\
\psi_{[21]_\rho}(q^2c)_2=&~\frac{1}{\sqrt{2}}(rgg-grg),\\
\psi_{[21]_\rho}(q^2c)_3=&~\frac{1}{\sqrt{2}}(rbr-brr),\\
\psi_{[21]_\rho}(q^2c)_4=&~\frac{1}{\sqrt{4}}(gbr+rng-brg-bgr),\\
\psi_{[21]_\rho}(q^2c)_5=&~\frac{1}{\sqrt{2}}(gbg-bgg),\\
\psi_{[21]_\rho}(q^2c)_6=&~\frac{1}{\sqrt{2}}(rbb-brb),\\
\psi_{[21]_\rho}(q^2c)_7=&~\frac{1}{\sqrt{2}}(gbb-bgb),\\
\psi_{[21]_\rho}(q^2c)_8=&~\frac{1}{\sqrt{12}}(2rgb-2grb-gbr\\
&+rbg-brg+bgr).
\end{split}
\end{equation}

The corresponding color octet states of the meson cluster are
\begin{equation}\label{colmes}
\begin{split}
&\psi_{[21]}(q\bar{c})_1=b\bar{r},\\
&\psi_{[21]}(q\bar{c})_2=b\bar{g},\\
&\psi_{[21]}(q\bar{c})_3=-g\bar{r},\\
&\psi_{[21]}(q\bar{c})_4=\frac{1}{\sqrt{2}}(r\bar{r}-g\bar{g}),\\
&\psi_{[21]}(q\bar{c})_5=r\bar{g},\\
&\psi_{[21]}(q\bar{c})_6=-g\bar{b},\\
&\psi_{[21]}(q\bar{c})_7=r\bar{b},\\
&\psi_{[21]}(q\bar{c})_8=\frac{1}{\sqrt{6}}(2b\bar{b}-r\bar{r}-g\bar{g}).
\end{split}
\end{equation}

\bibliography{biblio}

\end{document}